# Spin-valley Hall effects and pseudospin-1 system in transition-metal dichalcogenides: three band model approach

Phusit Nualpijit [a], Kitakorn Jatiyanon[b], Bumned Soodchomshom[c,*]

[a,c] Department of physics, Faculty of science, Kasetsart University, Bangkok 10900,Thailand

[b] Department of physics, Faculty of science, Rangsit university, Pathumthani 12000, Thailand

[a] School of *Integrated* Science (SIS), Kasetsart University, Bangkok 10900,Thailand

*Corresponding Author's E-mail: fscibns@ku.ac.th, bumned@hotmail.com

## Abstract

The energy spectra of transition metal dichalcogenides are primarily influenced by the $d_{z^2}, d_{xy}$ and $d_{x^2-y^2}$ orbitals. This results in a three-band model characterized by strong spin-orbit interaction. Investigating these bands using a projection into a two-band $\mathbf{k} \cdot \mathbf{p}$ model is adequate for exploring electronic and optical properties. The topological phases, spin-valley Hall effect, derived from spin-valley Chern numbers, are also examined. The inclusion of spin-orbit interaction breaks inversion symmetry while maintaining time-reversal symmetry. This leads to the non-zero spin-valley dependent Hall conductivities, pivotal for both the spin Hall and valley Hall effects. These phenomena could have applications beyond traditional electronics, potentially encoding quantum bits. In particular, electrons at high-symmetry points, namely the K and K' points, might behave as if they are in a pseudospin-1 system with spin angular momentum of $j=1$ and $m=0,\pm 1$, leading to triplet energy states. Such predictions could encourage super-Klein tunneling, which might offer insights into Majorana fermions, relevant for topological quantum computation.

**Keywords:** transition metal dichalcogenides, $MoS_2$, optical properties, pseudospin-1, quantum spin hall effect, quantum valley hall effect

# 1. Introduction

The spin-valley effect in transition metal dichalcogenides $MX_2$ (M=Mo,W; X=S, Se) is a key feature for electronic and optical properties, making it suitable for spintronic and valleytronic applications [1, 2]. Due to the broken inversion symmetry caused by spin-orbit coupling, two inequivalent valleys emerge in the momentum space within the Brillouin zone, leading to spin and valley-dependent electronic properties [2-4]. Magnetic doping offers a way to control both spin and valley polarization. For instance, substituting the S atom in $MoS_2$ with non-magnetic elements like H and B is predicted to result in a ferromagnetic phase [3, 5]. This arises from the coupling between the spin of the itinerant electron and the localized spin of the substituting atom, which is related to the fraction of dopant atoms and cation concentration [6]. Furthermore, $MoS_2$ undergoes a transition from semiconductor to half-metal when doped with H and B [5]. It's possible to manipulate the separation of the valence band maximum and the conduction band minimum through the proximity effect [7-9]. A substrate is a suitable candidate that can induce a large energy gap exceeding 300 meV. With the proximity effect, this gap can be adjusted by rotating the magnetization using a minor magnetic field [7, 10]. Recent discussions in Ref. [11]. have addressed predictions about the energy gap when rotating the angle of the in-plane magnetic field. The engineering of the energy band gap can be expressed analytically through the orientation of the in-plane magnetic field and the perpendicular electric field. Projections indicate that the ON and OFF states of the longitudinal current could be utilized for optical switches [11-13].

The valley degree of freedom in a hexagonal lattice plays a crucial role in momentum, characterized by a large separation between two valleys in momentum space. As a result, the intervalley scattering time is sufficiently long in the clean limit. Unlike spin, which is strongly affected by perturbations from impurities and thermal fluctuations that easily alter spin orientation, the valley is more stable. This makes it suitable for information storage and processing, with the main challenge being intervalley scattering [10, 14]. This potential paves the way for adopting the valley degree of freedom as a "quantum bit" in the emerging field of valleytronics. Theoretical predictions for controlling valley current are discussed in Ref. [15, 16]. In this context, the valley-locking state is the foundation of valley selection in response to circular polarization [15-17]. Recently, Sangeeta Sharma et al. proposed a method to induce and control valley current using a specially designed laser pulse [2]. The "honeycomb" pulse allows for complete control over the valley current by breaking inversion symmetry. This concept of breaking inversion symmetry is achieved by applying both linearly and circularly polarized pulses to the layer[2, 18]. Valley transport properties could be of interest in transition metal dichalcogenides since studies have revealed the emergence of spin and valley-dependent band structures [19-24], which might be influenced by external forces.

In this paper, we investigate the optical properties of transition metal dichalcogenides $MX_2$ using a 3-band tight-binding model derived from $d_{z^2}, d_{xy}$ and $d_{x^2-y^2}$ orbitals of an atom, taking spin-orbit coupling into account. The emergence of a spin-valley dependent band facilitates the Spin Hall and Valley Hall effects. This breaks the inversion symmetry but preserves time-reversal symmetry. We can simplify the optical problem using the Löwdin partitioning method, which projects the 3x3 matrix down to a solvable 2x2 matrix. In this work, we clarify that the Chern numbers, which include the quadratic term in the diagonal element of the Hamiltonian, exhibit dependency on the hopping parameter. We also elucidate why the spin and valley dependent Chern numbers aren't constant, yet remain consistent when combined in the Brillouin zone. Additionally, we demonstrate that the pseudospin-1 system at the K and K' points can be evaluated, and it behaves similarly to the spin-1 system. This insight paves the way for exploring the spin-1 system via two-dimensional Dirac fermions.

## 2. MODEL AND FORMALISM

In this section, we investigate the fundamental electronic properties of monolayer group-VIB transition metal dichalcogenides $MX_2$ ( $M = Mo, W$ and $X = S, Se, Te$ ), using tight-binding model. The theoretical studies indicates that $d$-orbitals from M atom especially $d_{z^2}, d_{xy}$ and $d_{x^2-y^2}$, make the primary contributions, while $s$-orbitals and $p$-orbitals could be neglected around band edges [19-21]. It is appropriate to construct a three-band tight-binding Hamiltonian for the low energy limit using the bases of $d_{z^2}, d_{xy}$ and $d_{x^2-y^2}$ orbitals. We can also use the magnetic quantum numbers to define basis vector states as of the form [22]

$$|0\rangle \equiv |d_{z^2}\rangle, \qquad |+2\rangle \equiv |d_{x^2-y^2}\rangle, \qquad |-2\rangle \equiv |d_{xy}\rangle \qquad (1)$$

The general form of the Hamiltonian describing the present system may be obtained as

$$H = \sum_{i,\mu} \left( E^a_{i\mu} a^\dagger_{i\mu} a_{i\mu} + E^b_{i\mu} b^\dagger_{i\mu} b_{i\mu} \right) + \sum_{\langle i,j \rangle, \mu\nu} \left( t_{ij,\mu\nu} a^\dagger_{i,\mu} b_{j,\nu} + h.c. \right)$$
$$+ \sum_{\langle\langle i,j \rangle\rangle, \mu\nu} \left( t'_{ij,\mu\nu} a^\dagger_{i,\mu} a_{j,\nu} + t'_{ij,\mu\nu} b^\dagger_{i,\mu} b_{i+j,\nu} \right) \qquad (2)$$

where $a^\dagger_i (a_i)$ are the creation (annihilation) operator of electron at sublattice $a$ with atomic site $i$ and orbital degree of freedom $\mu$. The on-site energy of the atom $a$ with orbital $\mu$ at the atomic site $i$ can be indicated by $\varepsilon^a_{i\mu}$. The hopping parameter $t_{ij,\mu\nu}$ has been sum over the nearest-neighbor atoms in which the nearest neighbor vectors of hexagonal lattice can be given in the real space by $\boldsymbol{\delta}_1 = a(0,-1)$, $\boldsymbol{\delta}_2 = (a/2)(\sqrt{3},1)$ and $\boldsymbol{\delta}_3 = (a/2)(-\sqrt{3},1)$. The next-nearest neighbor hopping in the same sublattice can be included with hopping parameter $t'_{ij,\mu\nu}$. For group-VIB transition metal dichalcogenides $MX_2$, $d$-orbitals from transition metal are mostly dominant. Thus, the nearest-neighbor hopping effect may be neglected. This becomes $3 \times 3$ matrices of Hamiltonian related to $d_{z^2}, d_{xy}$ and $d_{x^2-y^2}$ orbitals for next-nearest neighbor hopping [20]. The overlap integral for $M-M$ atoms can be evaluated by $t'_{ij,\mu\nu} \equiv \langle i,\mu | H | j,\nu \rangle$ where $\mu,\nu = 0, \pm 2$ related to the states of Eq.2. These parameters may be determined by fitting with the result given by the density functional theory in Ref. [20]. These hoping parameters may be defined as

$$t_0 \equiv \langle d_{z^2} | H | d_{z^2} \rangle, \qquad t_1 \equiv \langle d_{z^2} | H | d_{xy} \rangle, \qquad t_2 \equiv \langle d_{z^2} | H | d_{x^2-y^2} \rangle,$$
$$t_{11} \equiv \langle d_{xy} | H | d_{xy} \rangle, \qquad t_{12} \equiv \langle d_{xy} | H | d_{x^2-y^2} \rangle, \qquad t_{22} \equiv \langle d_{x^2-y^2} | H | d_{x^2-y^2} \rangle. \qquad (3)$$

They are negative or positive signs for forward and backward hopping due to symmetry of $d_{xy}$-orbital. Thus, the matrix element of overlap integral along the next-nearest neighbor vector $\mathbf{c}_1$ can be expressed by

$$H_{\pm c_1} = \begin{bmatrix} t_0 & \pm t_1 & t_2 \\ \mp t_1 & t_{11} & \pm t_{12} \\ t_2 & \mp t_{12} & t_{22} \end{bmatrix}. \tag{4}$$

On the other hand, the of overlap integral along $c_2$ and $c_3$ can be evaluated by symmetry transformation [20, 23] and the onsite energy in each orbital corresponding to the Hamiltonian in Eq.2 are $E_1, E_2$ and $E_3$, respectively. The large intrinsic spin-orbit coupling in $MX_2$ has been included due to large magnetic moment of transition metal atom, which may be defined as [20]

$$H_{so} = \frac{\Delta_{so}}{2} \begin{bmatrix} 0 & 0 & 0 \\ 0 & 0 & i \\ 0 & -i & 0 \end{bmatrix}. \tag{5}$$

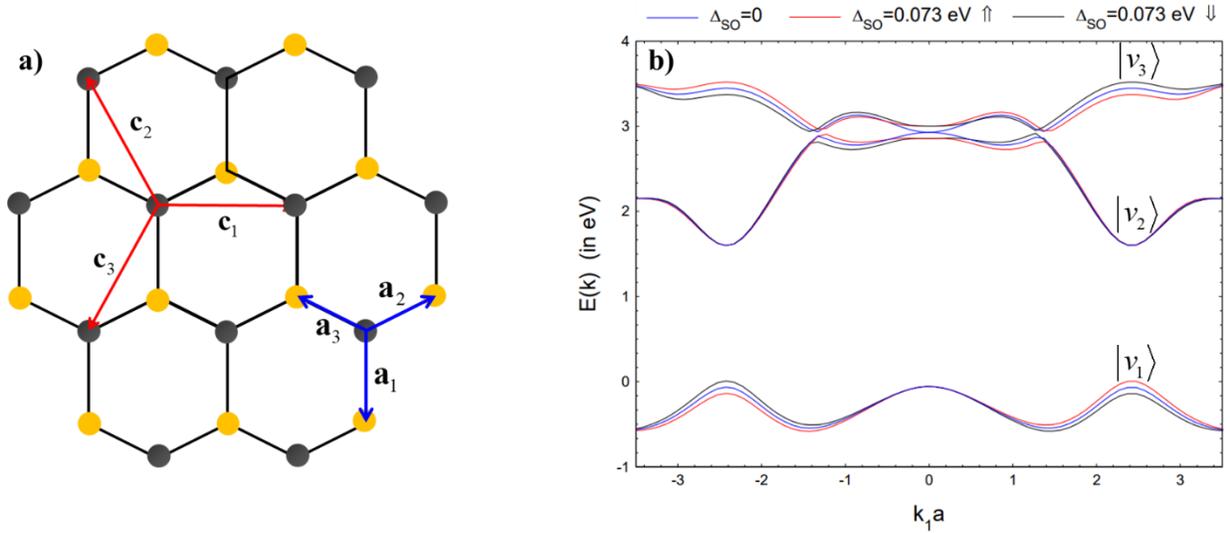

**Fig.1 a)** honeycomb lattice with the nearest neighbor vectors $a_i$ and next-nearest neighbor vectors $c_i$. **b)** The band structure of $MoS_2$ by tight-binding model with $d_{z^2}, d_{xy}$ and $d_{x^2-y^2}$ orbitals with and without spin-orbit interaction for spin up and spin down. The Eigenkets $|v_1\rangle, |v_2\rangle$ and $|v_3\rangle$ are related to the bottom band, middle band and top band, respectively. The parameters (in eV unit) are adopted from Ref.[20] which are $E_1 = 1.046$, $E_2 = 2.104$, $t_0 = -0.184$, $t_1 = 0.401$, $t_2 = 0.507$, $t_{11} = 0.218$, $t_{22} = 0.057$, $t_{12} = 0.338$, and $\Delta_{so} = 0.073$.

The band structure for $k_y = 0$, in the cases with and without spin orbit effect, is illustrated in Fig.1. This has deviated from case of $\Delta_{SO} = 0$ due to spin-orbit interaction affecting electron with both spin up and spin down. Investigating the electronic and optical properties by 3x3 matrix is intricate. However, the unitary operator $\mathbb{U}_0$ which transforms the nonorthogonal basis to an orthogonal one, can be evaluated at both K and K'-points [24]. Consequently, after unitary transformation, the Hamiltonian around K and K'-points [$\pm \mathbf{K} = (\pm 4\pi/(a3\sqrt{3}), 0)$] can be approximated by

$$H_\eta \equiv \mathbb{U}_0^\dagger H \mathbb{U}_0 \approx \begin{pmatrix} \varepsilon_1 & \alpha(k_1 - i\eta k_2) & \beta(k_1 + ik_2) \\ \alpha(k_1 + i\eta k_2) & \varepsilon_2 & \gamma(k_1 - i\eta k_2) \\ \beta(k_1 - i\eta k_2) & \gamma(k_1 + i\eta k_2) & \varepsilon_3 \end{pmatrix}, \qquad (6)$$

where 
$$\varepsilon_1 = E_2 - \frac{3}{2}t_{11} - \frac{3}{2}t_{22} - 3\sqrt{3}t_{12} + \eta\Delta_{SO},$$

$$\varepsilon_2 = E_1 - 3t_0,$$

$$\varepsilon_3 = E_2 - \frac{3}{2}t_{11} + \frac{3}{2}t_{22} + 3\sqrt{3}t_{12} - \eta\Delta_{SO},$$

$$\alpha = \eta\left(\frac{3\sqrt{6}}{4}t_1 + \frac{9\sqrt{2}}{4}t_2\right), \qquad \beta = \eta\frac{9}{4}(t_{22} - t_{11}), \qquad \gamma = \eta\left(-\frac{3\sqrt{6}}{4}t_1 + \frac{9\sqrt{2}}{4}t_2\right),$$

and $\eta = \pm 1$ for valley degree of freedom. To prevent the 3x3 matrix from being exactly unsolvable, the Löwdin partitioning method has been applied to the Hamiltonian in Eq. 6 to reduce its dimension to a 2x2 Hamiltonian [25]. The transformed Hamiltonian can be written as $H' \equiv e^{-\mathbb{O}} \mathbb{U}_0^\dagger H \mathbb{U}_0 e^{\mathbb{O}}$. The unitary operator $\mathbb{O}$ satisfies the condition $e^{-\mathbb{O}} H e^{\mathbb{O}} = H_{diag} + V + [H_{diag}, \mathbb{O}] + [V, \mathbb{O}] + \frac{1}{2}[[H_{diag}, \mathbb{O}], \mathbb{O}] + ...$ [26] which $H_{diag}$ is the diagonal block and $V$ has been treated as a perturbation. In this context, the bands associated with states $|v_1\rangle$ and $|v_2\rangle$ were selected for investigating the optical transport around nodal points, as the $|v_3\rangle$ band is distant enough to be disregarded. Hence, the 2x2 Hamiltonian can be expresses by

$$H_{tot}' = H_0' + V \approx \begin{pmatrix} \varepsilon_1 & \alpha(k_1 - i\eta k_2) \\ \alpha(k_1 + i\eta k_2) & \varepsilon_2 \end{pmatrix} + \frac{1}{2}\begin{pmatrix} V_{11} & V_{12} \\ V_{12}^* & V_{22} \end{pmatrix},$$

where

$$V_{11} = \frac{2\beta^2(k_1^2 + k_2^2)(\varepsilon_2 - \varepsilon_3)}{\alpha(k_1^2 + k_2^2) - (\varepsilon_1 - \varepsilon_3)(\varepsilon_2 - \varepsilon_3)}$$

$$V_{12} = \frac{\gamma\beta(k_1 + i\eta k_2)^2(\varepsilon_1 + \varepsilon_2 - 2\varepsilon_3)}{\alpha(k_1^2 + k_2^2) - (\varepsilon_1 - \varepsilon_3)(\varepsilon_2 - \varepsilon_3)}$$

$$V_{22} = \frac{2\gamma^2(k_1^2 + k_2^2)(\varepsilon_1 - \varepsilon_3)}{\alpha(k_1^2 + k_2^2) - (\varepsilon_1 - \varepsilon_3)(\varepsilon_2 - \varepsilon_3)}.$$

(7)

In this calculation, the spin orbit interaction related to Eq.6-7 are $\Delta_{SO} = \pm 0.073$ eV where + stands for spin up and − for spin down. The hopping parameters and onsite energies have been identified in the caption of Fig.1b adopted from Ref.[20]. The table 1 identifies the energy at K and K'-points which have been calculated from the Hamiltonian in Eq.7. The result is in agreement with the calculation from full band structure in Fig.1b. In this investigation, $|v_1\rangle$ is the valence band, while $|v_2\rangle$ is the conduction band.

|           | K-point ⇑ | K-point ⇓ | K'-point ⇑ | K'-point ⇓ |
|-----------|-----------|-----------|------------|------------|
| VBM       | 0.0082    | −0.1378   | −0.1378    | 0.0082     |
| CBM       | 1.5980    | 1.5980    | 1.5980     | 1.5980     |
| CBM − VBM | 1.5898    | 1.7358    | 1.7358     | 1.5898     |

**Table.1** The eigenenergies (in eV units) of electrons at the K and K' points for both spin up ⇑ and spin down ⇓ are identified by the valence band maximum (VBM) and conduction band minimum (CBM). The energy difference between CBM and VBM is used to specify the band gap.

# 3. OPTICAL CONDUCTIVITIES AND CHERN NUMBERS

The reaction of electrons under the electromagnetic field are expressed in terms of time fluctuation of dynamic variables. The energy operator of electric field $\mathbf{E}(t)$ perturbation can be written by $H_{int}'(t) = -\sum_i e_i \mathbf{x}_i \cdot \mathbf{E}(t)$ where $e_i$ is the charge of $i^{th}$ – particle with position $\mathbf{x}_i$. The response electric current in $\mu$ direction under electric field polarization $\nu$ can be investigated by Kubo formula [27]

$$\sigma_{\mu\nu}(\omega) = \int_0^\infty e^{-i\omega t} \int_0^\beta d\lambda \langle J_\nu(-i\hbar\lambda) J_\mu(t) \rangle, \qquad (8)$$

where $\beta = 1/(kT)$ and $J_\mu \equiv \sum e_i \dot{x}_{i\mu}$ is the current operator. To prevent the complication of analytical calculation, the $\mathbf{k} \cdot \mathbf{p}$ Hamiltonian in Eq.7 has been considered to investigate the electronics and optical behaviors of electrons around the nodal points. The $\mathbf{k} \cdot \mathbf{p}$ Hamiltonian can be written in terms of Pauli matrices $\sigma_i$ as

$$H_0' \approx h_0 \sigma_0 + h_1 \sigma_1 + h_2 \sigma_2 + h_3 \sigma_3, \qquad (9)$$

where $h_0 = \frac{1}{2}(\varepsilon_1 + \varepsilon_2)$, $h_1 = \alpha k_1$, $h_2 = \alpha k_2$ and $h_3 = \frac{1}{2}(\varepsilon_1 - \varepsilon_2) \equiv \frac{1}{2}\Delta$. The corresponding eigenvalues and eigenvectors are respectively given as

$$E_\pm = h_0 \pm E_0 = h_0 \pm \sqrt{h_1^2 + h_2^2 + h_3^2}, \qquad (10)$$

and $\quad |\psi_\pm\rangle = \frac{1}{\sqrt{2E_0(E_0 \pm h_3)}}(\pm E_0 + h_3, h_1 + ih_2)$. (11)

The calculation can be implemented analytically for both real part $\sigma_{xx}'(\omega)$ and imaginary part $\sigma_{xx}''(\omega)$ of longitudinal conductivity, given as the same as in Ref.[9, 28]. These quantities read

$$\sigma_{xx}'(\omega) = \frac{\pi e^2}{8h}\left[1+\left(\frac{\Delta}{\omega}\right)^2\right]\Theta(\omega-|\Delta|)\ ,\qquad(12)$$

$$\sigma_{xx}''(\omega) = \frac{\pi e^2}{8h}\left[\frac{2\Delta}{\pi\omega} - \frac{2\Delta^2}{\pi\omega\varepsilon_{cut}} + \frac{1}{\pi}\left(1+\frac{\Delta^2}{\omega^2}\right)\left(\ln\left|\frac{\varepsilon_{cut}+\omega}{\varepsilon_{cut}-\omega}\right| + \ln\left|\frac{\Delta-\omega}{\Delta+\omega}\right|\right)\right],\qquad(13)$$

where $\sigma_0 = \frac{\pi e^2}{8h}$ is the conductivity per spin and valley degree of freedom. $\varepsilon_{cut} \gg |\Delta|$ is the cutoff energy related to the band structure of energy states $\{|v_1\rangle, |v_2\rangle\}$ given in Fig.1, and $\omega$ is the photon energy. Furthermore, the two transverse conductivities may be achieved as

$$\sigma_{xy}'(\omega) = \eta\frac{\pi e^2}{8h}\left(\frac{2\Delta}{\pi\omega}\right)\left[\ln\left|\frac{\omega+\varepsilon_{cut}}{\omega+\Delta}\right| - \ln\left|\frac{\omega-\varepsilon_{cut}}{\omega-\Delta}\right|\right],\qquad(14)$$

$$\sigma_{xy}''(\omega) = \eta\frac{\pi e^2}{8h}\left(\frac{2(\varepsilon_1-\varepsilon_2)}{\omega}\right)\Theta(\omega-|\Delta|).\qquad(15)$$

The Berry connection may be defined as $A_\mu = -i \sum_i \langle \psi_k | \frac{\partial}{\partial k_\mu} | \psi_k \rangle$, which is operated on the occupied state $|\psi_k\rangle$. It is available to calculate DC transverse conductivity by using Chern number [29]

$$C = \frac{1}{2\pi} \oint_{\mathbb{C}} d\mathbf{k} \cdot \mathbf{A}(\mathbf{k}), \tag{16}$$

By the line integral along the closed loop $\mathbb{C}$. The relation between transverse conductivity and Chern number can be expressed by

$$\sigma_{xy}(0) = \frac{e^2}{h} C. \tag{17}$$

The calculation of Chern number around the nodal points with the Hamiltonian in Eq.9 can be expressed in terms of valley degree of freedom and sign of mass term as

$$C_\eta = -\frac{\eta}{2} \text{sgn}(\Delta), \tag{18}$$

in which the results agree with the approximation for $\omega \to 0$ of Eq. 14, by using the series expansion of $\ln\left(\frac{1+x}{1-x}\right) \approx 2x$ for $x \ll 1$.

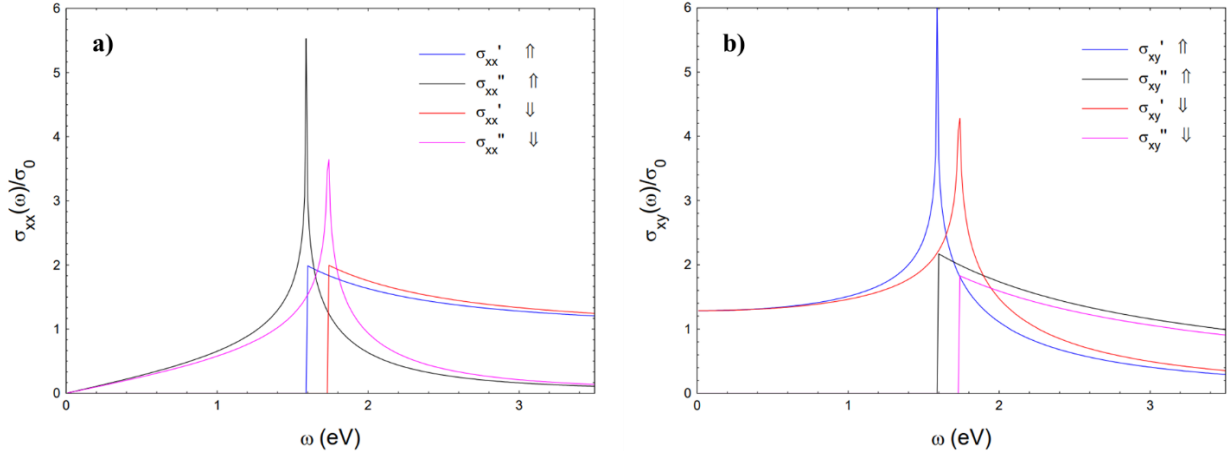

**Fig.2 a)** The real part $\sigma_{xx}{}'$ and imaginary part $\sigma_{xx}{}''$ of longitudinal conductivity around the vicinity of K-point for both spin up $\Uparrow$ and down $\Downarrow$. **b)** The transverse conductivity.

## 4. Pseudospin-1 system

Recalling Eq.6, the linearized $\mathbf{k}\cdot\mathbf{p}$ Hamiltonian around K and K' points in this system would take the form of a host of pseudospin-1, obtained as

$$H_K \approx \alpha k_1 J_1 + \alpha k_2 J_2 + \Delta J_3, \qquad (19)$$

where $\alpha$, $\Delta$ are adopted from Eq.6. The pseudospin-1 operators read

$$J_1 = \frac{1}{\sqrt{2}}\begin{pmatrix} 0 & 1 & 0 \\ 1 & 0 & 1 \\ 0 & 1 & 0 \end{pmatrix}, \qquad J_2 = \frac{1}{\sqrt{2}}\begin{pmatrix} 0 & -i & 0 \\ i & 0 & -i \\ 0 & i & 0 \end{pmatrix}, \qquad J_3 = \begin{pmatrix} 1 & 0 & 0 \\ 0 & 0 & 0 \\ 0 & 0 & -1 \end{pmatrix}.$$

These operators are satisfied the commutation relation $[J_i, J_j] = i\varepsilon_{ijk} J_k$ with cyclic permutation $\varepsilon_{ijk}$. The wave states of speudospin-1 system at the K and K' points. By substituting $k_1 \to 0$, $k_2 \to 0$ which are spin-1 eigenvectors and eigenvalues are respectively given as

$$|v_{-1}\rangle = \begin{pmatrix} 0 \\ 0 \\ 1 \end{pmatrix} \qquad |v_0\rangle = \begin{pmatrix} 0 \\ 1 \\ 0 \end{pmatrix} \qquad |v_1\rangle = \begin{pmatrix} 1 \\ 0 \\ 0 \end{pmatrix}, \tag{20}$$

and $\varepsilon_s = m\Delta$ \hfill (21)

where $m = 0, \pm 1$ being pseudo spin magnetic numbers. These can be easily shows that the eigenvalue problem of $J^2|j,m\rangle = j(j+1)|j,m\rangle$ and $J_3|j,m\rangle = m|j,m\rangle$ with $j=1$ for pseudospin-1 system [30]. This suggests the eigenstates to be the simultaneous eigenkets related to quantum number $j,m$. Furthermore, the raising and lowering operators for changing quantum number $m$ with a fixing $j$ can be expressed by $J_\pm = J_1 \pm iJ_2$. The Hamiltonian behaves like precession of charge particle under pseudo uniform magnetic field along $J_3$ component. The time-dependent eigenstates can be evaluated by applying time-evolution operator

$$|\psi(t)\rangle \equiv \mathbb{U}(t)|\psi(t=0)\rangle, \tag{22}$$

where $\mathbb{U}(t) = e^{-iHt}$ and $|\psi(t=0)\rangle = a|v_1\rangle + b|v_2\rangle + c|v_3\rangle$ with $|a|^2 + |b|^2 + |c|^2 = 1$. In this calculation, we choose $a = \cos\frac{\theta}{2}$, $b = \sin\frac{\theta}{2}\cos\phi$, and $c = \sin\frac{\theta}{2}\sin\phi$ with zenith angle $\theta$ and azimuth angle $\phi$. The expectation of pseudospin-1 operator as a function of time reads

$$\langle\psi(t)|J_1|\psi(t)\rangle = \sqrt{2}\cos(\Delta t)\sin\left(\frac{\theta}{2}\right)\sin\phi\left[\cos\left(\frac{\theta}{2}\right)+\cos\phi\sin\left(\frac{\theta}{2}\right)\right] \quad (23)$$

$$\langle\psi(t)|J_2|\psi(t)\rangle = \sqrt{2}\sin(\Delta t)\sin\left(\frac{\theta}{2}\right)\sin\phi\left[\cos\left(\frac{\theta}{2}\right)+\cos\phi\sin\left(\frac{\theta}{2}\right)\right] \quad (24)$$

$$\langle\psi(t)|J_3|\psi(t)\rangle = \sin^2\left(\frac{\theta}{2}\right)\cos^2\phi - \cos^2\left(\frac{\theta}{2}\right) \quad (25)$$

The results show that $\Delta$ can be interpreted as same as Larmor frequency [31].

## 5. RESULTS AND DISCUSSIONS

In this investigation, the electron especially from $d_{z^2}, d_{xy}$ and $d_{x^2-y^2}$ orbitals make the main contributions, while $s$-orbitals and $p$-orbitals have been neglected [19-21]. The transformation of Hamiltonian by diagonalization around K and K' points allows the system to achieve orthogonal states. Furthermore, the projection operator may offer an appropriate solution to address the unsolvable problem of 3x3 matrix for optical properties.

The two bands of states $|v_1\rangle$ and $|v_2\rangle$ from Fig.1 are studied in the vicinity of K and K' points. Meanwhile the band $|v_3\rangle$ has been neglected due to its long distance from nodal points. The vertical transition of electrons from $|v_1\rangle$ to $|v_2\rangle$ induces the optical current for each spin and valley degree of freedom, calculated by the Kubo formula as outline in Eq.12-13. The isotropic longitudinal conductivities are frequency-dependent because of gap opening [28] and are related to the effective onsite energy $\varepsilon_1, \varepsilon_2$ of $\mathbf{k}\cdot\mathbf{p}$ model. The results indicate that $\sigma_{xx}^{K\Uparrow} = \sigma_{xx}^{K'\Downarrow}$ and $\sigma_{xx}^{K\Downarrow} = \sigma_{xx}^{K'\Uparrow}$, which allow for the investigation of the spin and valley current. The charge conductivity can be defined by $\sigma_{xx}^q = \sigma_{xx}^{K\Uparrow} + \sigma_{xx}^{K\Downarrow} + \sigma_{xx}^{K'\Uparrow} + \sigma_{xx}^{K'\Downarrow}$. This can be applied as a frequency-dependent switch: it operates in the infrared regime with wavelength $714-780$ nm at a lower current state and in the visible regime of wavelength less than 714nm at a higher current (twice the magnitude) as discussed in Ref.[16]. The spin-valley conductivity, which can be respectively defined by expression $\sigma_{xx}^{v} = \left(\sigma_{xx}^{K\Uparrow} + \sigma_{xx}^{K\Downarrow}\right) - \left(\sigma_{xx}^{K'\Uparrow} + \sigma_{xx}^{K'\Downarrow}\right)$ and $\sigma_{xx}^{s} = \left(\sigma_{xx}^{K\Uparrow} + \sigma_{xx}^{K'\Uparrow}\right) - \left(\sigma_{xx}^{K\Downarrow} + \sigma_{xx}^{K'\Downarrow}\right)$, becomes zero. The frequency-dependent longitudinal conductivities are illustrated in Fig. 2a, in which the threshold frequencies corelate with energy gap presented in Table.1. The optical absorption can be calculated using Fermi's golden rule, which indicates the number of transitions of an electron from state $|v_i\rangle$ to $|v_f\rangle$ per unit of time per unit volume [32]. The absorption responses rapidly when the photon energy approaches the band gap, exhibiting behavior similar to that of $\sigma_{xx}$. This implies that the optical absorption provides insight into the magnitude of the band gap. This model might also be relevant for semiconducting transition metal carbides (MXenes) that possess a hexagonal structure with a pronounced magnetic moment [33, 34]. Surface functionalization with $F$, $O$, -OH group on MXenes affect the spin orientation, which can be observed through optical current [33, 34]. Given its significant electromagnetic absorption, the $MX_2$-like structure can serve as an energy absorber and a tool for gas detecting [33, 35].

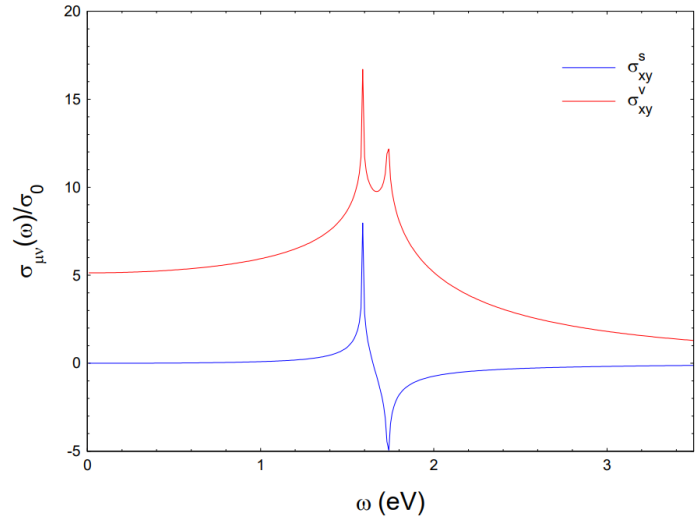

**Fig.3** The plot of real part spin and valley hall conductivity. The singularities as peak and dip curves are related to the energy gap indicated in Table.1.

The transverse optical response - the charge, spin and valley conductivities - are defined by the same way as in the longitudinal conductivity. The calculations in

Eq.14-15 reveal that $\sigma_{xy}^{K\Uparrow}=-\sigma_{xy}^{K'\Downarrow}$ and $\sigma_{xy}^{K\Downarrow}=-\sigma_{xy}^{K'\Uparrow}$. These exhibit behavior analogous to that observed in silicene with a non-uniform gap and consistent spin polarization [36]. Both theoretical and experimental evidence [11, 37] indicates that the substantial magnitude of Hall conductivity originates from the pronounced Berry curvature identified in narrower band gaps. As considered from Fig. 2b, while the real valley current remains consistent regardless of photon energy, the spin current alters its direction when $\sigma'^{K\Uparrow}_{xy}=\sigma'^{K\Downarrow}_{xy}$, as discussed in Ref.[15]. The spin orbit coupling breaks inversion symmetry, which is the key behind creation of spin and valley hall effect as shown in Fig.3. The spin Hall effect can be detected within a narrowly permissible range of photon energy, centered around the magnitude of the band gap. Conversely, the valley Hall effect remains relatively constant at low energy levels and exhibits high magnitude around the energy gap, similar to the spin Hall effect. This suggests that $MX_2$ experiences a high signal-to-noise ratio, spanning from the infrared to the visible regime. Regrettably, both the spin Hall and valley Hall effects are absent at the high photon energy of $\omega \gtrsim 3.5$ eV. With this structure, it might be feasible to achieve non-zero charge conductivity, unassociated with the Landau level and connected to the sign of mass terms [38], through the proximity effect of antiferromagnetic coating on the bilayer of $MX_2$. The energy gap can be conveniently modulated using a perpendicular electric field, leading to alterations in topological phases [8, 39]. The proximity effect breaks time-reversal symmetry, resulting in a non-uniform gap, which holds promise for spintronic and valleytronic devices [15, 40].

The Chern number in Eq.16 may be manifested as a surface integral of Berry curvature, $\Omega_{\mathbf{k}}=\nabla\times\mathbf{A}(\mathbf{k})$, by using Stroke's theorem

$$C=\frac{1}{2\pi}\oint_{\mathbb{C}} d\mathbf{k}\cdot\mathbf{A}(\mathbf{k}) \longrightarrow \frac{1}{2\pi}\int \left(\hat{z}d^2\mathbf{k}\right)\cdot(\nabla\times\mathbf{A}(\mathbf{k})),$$

which is expected to be a constant value when integrating over the whole Brillouin zone [41]. To study the analytical calculation of the Chern number around the vicinity of the K and K' points, the $k^2$ approximation in $h_3$ of the Hamiltonian in

Eq.9 should be included, because the magnitude of the coefficient is comparable to $\Delta$. The $k^2$ approximation in $h_1$ and $h_2$ is small by comparing with $\alpha$. Thus, these terms are negligible. The modified Hamiltonian around two nodal points can be therefore expressed by

$$H_{mod,\eta}' \approx \begin{pmatrix} \frac{1}{2}\Delta + \rho(k_1^2 + k_2^2) & \alpha(k_1 - i\eta k_2) \\ \alpha(k_1 - i\eta k_2) & -\left(\frac{1}{2}\Delta + \rho(k_1^2 + k_2^2)\right) \end{pmatrix}. \tag{20}$$

The parameters $\rho = \frac{9}{8}(t_{11} - \eta 2\sqrt{3}t_{12} + t_{22}) - \frac{9}{4}t_0$, $\alpha$ and $\Delta$ have been defined in Eq.9. Thus, the Chern number around the K and K' points can be expressed analytically as

$$C_\eta \approx \eta \frac{\Delta + 2\rho\Lambda^2}{2\sqrt{\Delta^2 + 4\rho^2\Lambda^4 + 4\Lambda^2 + 4\rho\Delta\Lambda^2}} - \eta \frac{\Delta}{2|\Delta|} \tag{21}$$

where $\Lambda$ is the cut-off energy. When we have $\rho = 0$ and $\Lambda \gg \Delta$, the result becomes the Chern number given in Eq.18, which is valley dependent. Interestingly, as the non-zero of $k^2$ is included in diagonal element of the Hamiltonian indicated by coefficient $\rho$, the results of the Chern number can be illustrated in Fig.4. The results state that the Chern number depends on the sign of mass term $\Delta$ as discussed in Ref. [38] and changes with parameter $\rho$. It can be seen that the Chern number approaches unity or zero, when the magnitude of $\rho$ is comparable to $\Delta$. The analytical results also show that it depends on the cut-off energy. This means that the cut-off energy should be large enough until $\Delta$ is insignificant. Furthermore, in the opposite valley with the same spin, the Chern number is found to be opposite $C_K = -C_{K'}$. This is in agreement with the sign of mass term indicated by $\Delta$, which has the same sign in this material. The variation of the Chern number is also observed in strained graphene, in which the Rashba spin orbit coupling is included in diagonal terms by varying the strength of interaction [42]. Even though the Chern number change with some parameter for each valley degree of freedom, it is still invariant within the Brillouin zone [18] by combining of two valley degree of freedom. This is consistent with the discussion of Haldane in the transverse conductivity [38].

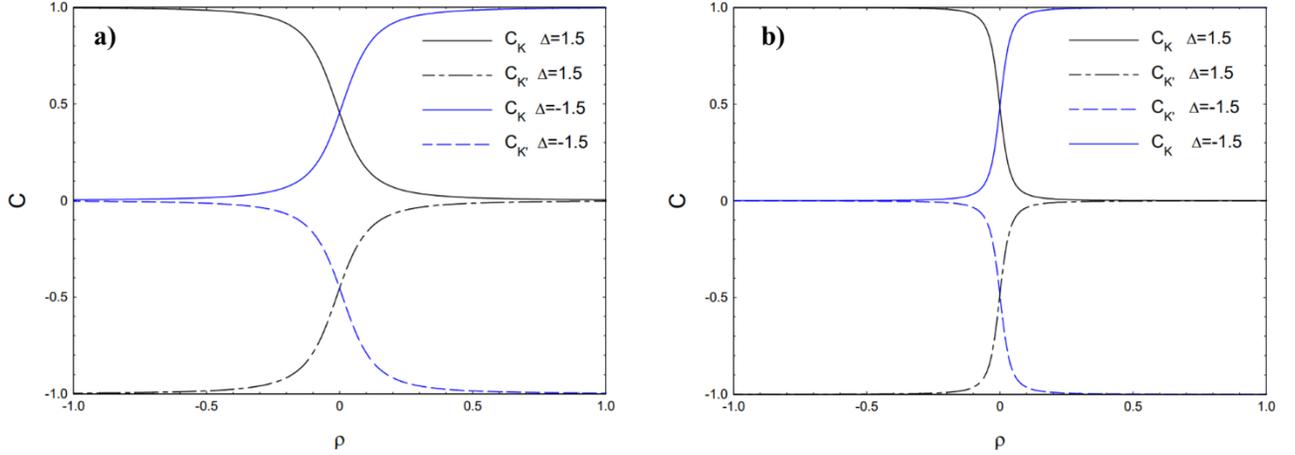

**Fig.4** a) The analytical calculation of Chern number as a function of $\rho$ with $\Delta$=1.5 eV for cutoff wave vector $k_{cut} \approx 5\Delta$ and b) $k_{cut} \approx 15\Delta$. The momentum-dependent term in diagonal elements of the Hamiltonian induce variation of Chern number, while the summation of the Chern numbers of the 2 valleys is still constant.

The pseudospin-1 system, novel fermion, may be realized in transition metal dichalcogenides $MX_2$, which are hosted by strong contribution of $d_{z^2}, d_{xy}$ and $d_{x^2-y^2}$ orbitals. It can be seen that the effect can be investigated at K and K' points depending on the mass term $\Delta$ as given in Eq.20-21. The Klein tunnelling of pseudospin-1 system with square potential barrier exhibits the complete transmission for all incident angle when the energy of electron is one-half of barrier potential [43]. This paves the way to make an electronics device with low energy loss. Recall Eq.23-25, the expectation value of pseudospin in $J_1$ and $J_2$ components disappear when $\theta = 0$ or $\phi = 0$. This means that the expectation value of $J_3$ components is in stationary state without oscillating with time. When the energy gap almost approaches zero $\Delta \to 0$, the approximation shows that the expectation value of $J_2$ components goes to zero and $J_1$ becomes constant. This means that the pseudospin is localized with fixed $\theta$ and $\phi$. Furthermore, the frequency of pseudospin precession can be controlled by energy gap $\Delta$ depending on the hopping of substance. This may lead to the mechanical force controlling the electronic behavior of pseudospin-1 system. This phenomena may be also observed in a Kagome lattice with flat band and Dirac cone [44, 45]. The Hamiltonian can be modified by diagonalization process around Dirac point which is the pseudospin-1 generated. This expression of pseudospin-1 may provide an in-deep comprehension and may begin studying superconductivity in $MX_2$-like structure, Kagome lattice, and MXenes structure [43, 46, 47].

## 6. SUMMARY AND CONCLUSION

We have conducted an analytical investigation into the electronic properties and optical conductivities of transition metal dichalcogenides $MX_2$, taking into account the spin-orbit interaction. The three-band tight-binding model encompasses the $d_{z^2}, d_{xy}$ and $d_{x^2-y^2}$ orbitals. Within this model, the Fermi level is posited to be at the pinnacle of the lowest band. Consequently, projecting these three bands onto the lowest and middle bands suffices for probing their electronic and optical characteristics. The inclusion of spin-orbit interaction breaks inversion symmetry while maintaining time-reversal symmetry. This leads to the manifestation of non-zero spin-valley dependent Hall conductivities, pivotal for both the spin Hall and valley Hall effects. Spintronics and valleytronics emerge as prospective platforms transcending traditional electronics, particularly for the encoding of quantum bits. However, a caveat is that spin excitations are susceptible to thermal fluctuations and impurity perturbations, rendering valley excitations as more stable counterparts for quantum bit representation. This stability is primarily challenged by intervalley scattering alone [2]. Research into the pseudospin-1 system, when modeled with three bands at the high symmetry points K and K' becomes feasible in the context of $MX_2$. The eigenspaces corresponding to the Hamiltonian at these symmetry junctures resemble an angular momentum system with $j=1$ and $m=0,\pm 1$, as elaborated in Section 4. Moreover, one can discern angular-independent super Klein tunneling across a static barrier [48, 49], which offers insights into the behavior of massive pseudospin-1 particles. These particles could potentially be the foundation for topological quantum computation using Majorana fermions [50].

# ACKNOWLEDGEMENT

This project is funded by National Research Council of Thailand (NRCT): NRCT5-RSA63002-15.